\documentclass[11pt]{article} 

\usepackage[utf8]{inputenc} 
\usepackage{geometry} 
\usepackage{graphicx} 
\usepackage{booktabs} 
\usepackage{array} 
\usepackage{amsmath}
\usepackage{verbatim} 
\usepackage{hyperref}

\title{Electoral David vs Goliath: How does the Spatial Concentration of Electors affect District-based Elections?}
\author{Adway Mitra\\
Indian Institute of Technology, Kharagpur\\}

\begin{document}
\maketitle

\abstract
Many democratic countries use district-based elections where there is a ``seat" for each district in the governing body such as parliament or assembly. In each district, the party whose candidate gets the maximum number of votes is declared the winner of the corresponding seat. The result of the election is decided based on the number of seats won this way by the different parties. The electors (or voters) are assigned to different districts based on their residence, and each elector can vote only in the district in which they are assigned. Thus, locations of the electors and boundaries of the districts may severely affect the election result even if the proportion of popular support (number of electors) of different parties remains unchanged. In this setting, it is also possible that a party with less number of supporters overall, may end up winning more seats if their supporters are suitably distributed. This has led to significant amount of research on the topic of ``gerrymandering", i.e. how districts may be redrawn or electors may be moved to ensure maximum seats for a particular party. In this paper, we frame the spatial distribution of electors in a probabilistic setting, and analyze different models to capture the intra-district polarization of electors in favour of a party, or the spatial concentration of supporters of different parties. Our models are inspired by elections in India, where supporters of different parties tend to be concentrated, turning various districts into strongholds of certain parties. We show with extensive simulations that our model can capture different statistical properties of real elections held in India. For this purpose, we frame parameter estimation problems to fit our models to the observed election results. Since analytical calculation of the likelihood functions are infeasible for our complex models, we make use of Likelihood-free Inference methods under the framework of Approximate Bayesian Computation (ABC). 
Since this approach is highly time-consuming, we explore how supervised regression using Logistic Regression or Deep Neural Networks can be used to speed it up. We also explore how the election results can change drastically by varying the spatial distributions of the voters, even when the proportions of popular support of the parties remain constant. 
%We also study which elections in India, UK and Israel have seen strong and weak concentrations of voters, and which regions have strong and weak polarization.

\section{Introduction}
Elections are conducted by almost all democratic countries to choose representatives for governing bodies, such as parliaments. While there are many kinds of electoral systems, a common form is the district-based system in which the country is spatially divided into a number of regions called districts (or constituencies). There is a seat in the governing body corresponding to each district. So the residents of each district elect a representative from a set of candidates, according to any voting rule. The district representatives thus elected to the governing body, may then form the government by forming alliances. However, in many countries, the district-level candidates are actually representatives of political parties, and when the electors cast their votes, they may do so in favour of the parties, rather than the individual candidates.

The election results are understood in terms of the number of seats won by different parties, rather than the total number of votes obtained by them. If the relative popularity of the different parties is spatially homogeneous across all the districts, then we can expect the most popular party winning all the seats. But this is very rarely the case. One reason for this may be the individual popularity of certain candidates, who are elected from their respective districts irrespective of their party affiliations. A more complex reason is the spatial variation of demography across the country, coupled with the fact that the popularity of different parties varies with demography~\cite{b}. Demography varies spatially as people usually prefer to choose residences based on social identities, such as race, religion, language, caste, profession and economic status. This process is sometimes called ``ghettoization", where people with similar social identities huddle together in pockets~\cite{d,e}. Such ghettoization plays a very important role in district-based elections if different political parties represent the interests of different social groups. Even if a political party is not popular overall, it can win a few seats if its supporters are densely concentrated in a small number of districts, which forms strongholds of the party. On the other hand, a party which is overall quite popular, may end up failing to win many seats if its supporters are spread all over without concentration. Additionally, even within each district, electors often vote not according to individual choice but following the advice of community leaders, and other local factors~\cite{c}, which may result in ``polarization" of voters in favour of one or two parties inside each district.

These spatial effects make district-based elections very complex for both politicians and psephologists, for strategy-formation and forecasting/interpretation of results respectively. Often the relative popularity of parties can be understood through surveys. But failure to understand the spatial distribution of such popularity limits the ability to estimate the number of seats that may be won by the parties. Also, the spatial distribution of popularity is not static over time as people often migrate and relocate, old electors pass away and young electors get registered. Finally, since the process of partitioning the country into districts is exogenous to the election process, the robustness and comprehensiveness of the results are also questionable. In fact, many countries have the malpractice of ``gerrymandering" in which parties having executive powers try to redefine the districts with the aim of maximizing their seats in upcoming elections, even without improving their relative popularity. The number of seats won by the parties should be (almost) invariant to the way in which districts are drawn up, if the results are to be considered as fair and robust. 

The aim of this paper is to build a framework to explore and understand the influence of the spatial effects mentioned above on the results of elections. Given the relative popularity of different political parties, we seek to estimate the number of seats that they may win, while the partitioning of the electors on the basis of districts is considered as a random process. We consider different probabilistic models for this purpose, which are capable of capturing the phenomena like ghettoization and local polarization as discussed above. We demonstrate our results on synthetic data, as well as on real data based on elections in India, which has one of the most complex electoral processes in the world due to the presence of many political parties and a highly heterogeneous and layered society where social identities are deeply interlinked with politics, resulting in spatial heterogeneity in election results~\cite{a}. To fit our models to observed electoral data from India, we explore the parameter estimation problem. However, the proposed models are ``generative-only": from which samples can be drawn but analytical computation of likelihood function is infeasible. So we take the help of likelihood-free inference techniques under the realm of Approximate Bayesian Computation (ABC). For this purpose, we design summary statistics of the election results which are both useful for the ABC techniques but also useful to understand election results. For one of these models, there is only one scalar parameter so we can compute its posterior distribution using simple ABC Rejection Algorithm. However for other models the parameter space is much larger and so the rejection algorithm is too inefficient. Hence we modify the ABC Rejection algorithm to make a more focused search. We also explore the usefulness of regression algorithms based on logistic regression and neural networks using our sufficient statistics to find a rough estimate of the parameters, from which the rejection approach can take over.

The paper is organized as follows. In Section 2 we review the literature on spatial issues in district-based elections (especially gerrymandering), and Likelihood-free Inference techniques. In Section 3 we discuss our probabilistic models for the spatial distribution of electors. Section 4 explores parameter estimation for the models using the ABC framework, while in Section 5 we explore regression techniques for this purpose. In Section 6 we show the simulation results of our models on synthetic data, while in Section 7 we carry out an analysis of recent elections in India using the proposed framework.

\section{Related Works}
\subsection{Elections and Gerrymandering}
A significant amount of literature exists in the domain of computational social science regarding district-based elections. However, most of these works concentrate on the topic of gerrymandering - a process of altering the districts to favour one particular party. The work \cite{3} introduces and examines different algorithms of manipulating elections, but using a different voting scheme than the one considered here. The electors are partitioned into two or more groups so that a candidate is selected from each partition, followed by another round of election among the selected ones, and algorithms are considered for creating such partitions under various constraints so that a particular candidate can be guaranteed to win. The works\cite{1,6} consider a setting where a subset of an initial set of districts may be retained and the rest merged, and suggest heuristic algorithms to maximize the number of districts won by a particular party. The work~\cite{2} introduces geometric constraints such as contiguity in defining districts, and explores the relationship between vote share and number of districts won in a two-party setting. This is the only work to the best of our knowledge which explores the issue of spatial distribution of supporters of two parties, though in a very restricted setting. In the same spirit, the work~\cite{4} considers the redrawing of districts with the aim of making the election more ``competitive", i.e. improving the number of seats of the less popular parties, by utilizing the geometric heterogeneity of vote share. On the other hand, the work~\cite{5} considers a game-theoretic setting in which the electors are rational agents who can relocate to another district, provided it improves the chances of their preferred outcome. 

\subsection{Approximate Bayesian Computation and Likelihood-free Inference}
Every stochastic process involves one or more parameters related to conditional or joint probability distributions. Fitting such models to observations requires us to estimate these parameters. We may be interested in either point estimates or a posterior distribution over the parameter space, depending on the application. However, well-known parameter estimation approaches require us to evaluate the likelihood function, i.e. the probability that the model, under a given parameter setting, will be able to generate the observed data. However if the stochastic process is complicated, then analytically calculating this probability may not be tractable. In such a situation, we try to carry out \emph{likelihood-free inference}, by either making an approximation of the likelihood function or by avoiding the Bayesian equation and directly estimating  the posterior distribution by drawing samples. This approach is known as \emph{Approximate Bayesian Computing (ABC)}.

One of the earliest known application of likelihood-free inference was in the domain of ecology~\cite{abc,abcrejection}, where the ABC rejection algorithm was introduced. This algorithm proceeds by drawing parameter values from a prior distribution, using that value to run the process simulation, and accept it only if the simulated outcome was close enough to the observed values. Using the accepted values of the parameters, a posterior distribution over the parameter space, conditioned on the observations, can be calculated. Because comparing the outcomes in full details is often difficult and usually futile, so~\cite{summary} suggested that only some \emph{summary statistics} of the outcomes and observations can be computed and compared. Such summary statistics may be provided by experts of the process, but may also be estimate in a data-driven way using neural networks~\cite{neural2}.

One major problem of this particular approach is that most of the samples will be rejected, so that the algorithm will have to run very long till sufficient number of samples have been collected. Additionally, if each simulation is time-consuming, then this approach may become infeasible. ~\cite{slam} improvises the algorithm to navigate the parameter space more smartly, so that we can move rapidly towards the acceptable parameters. Another body of works tries to circumvent the need to simulate for every parameter sample, and instead tries to predict whether the sample will be acceptable or not by training a classifier such as logistic regression~\cite{logistic} or by constructing a \emph{synthetic likelihood}\cite{bayesopt} for the summary statistics mentioned above, and accepting/rejecting each sample on the basis of the ratios of such likelihood~\cite{lratio}. Another approach look to substitute the rejection process altogether, and replace it with regression to map the each observation to a parameter value, using something like a neural network~\cite{emulator,neural1,neural2}. The data needed to train such neural networks can be obtained by running the simulation for a range of parameter values.

\section{Model and Analysis}
In this section, we first define the notations of our setting, and then proceed to discuss a series of models for spatial distribution of electors.

\subsection{Notation}
Let the total number of districts be $S$. There is one seat in the parliament corresponding to each district. The total number of electors is $N$, and each elector must register themselves in one district. Let $Z_i$ denote the district in which elector $i$ registers themselves. But each district can have a fixed number of electors, denoted by $\{n_1,n_2\dots,n_S\}$. Clearly,  $n_1+n_2+\dots+n_S=N$. Now, there are $K$ political parties, and the numbers of their supporters are $\{v_1,v_2,\dots,v_K\}$, such that $v_1+v_2+\dots+v_K=N$. The relative vote shares of these parties can be considered as a $K$-dimensional discrete distribution, $\theta$. In the subsequent analyses, we consider all the above quantities except $Z_i$ to be fixed and known, unless otherwise stated.

In the electoral setting, let the number of votes polled by the different parties at any district $s$ be denoted by $\{V_{s1},V_{s2},\dots,V_{sK}\}$. Clearly, $\sum_{k=1}^KV_{sk}=n_s$ and $\sum_{s=1}^SV_{sk}=v_k$. In any district $s$, the winner $W_s$ is that party which receives the highest number of votes in that district, i.e. $W_s = argmax_{k}(V_{s1},V_{s2},\dots,V_{sK})$.  In each district, the ``winning margin" $P_s$ is the fraction of votes won by the winning party, i.e. $P_s = \frac{V_{sW_s}}{n_s}$. The number of seats $M_k$ won by any party $k$ is the number of districts where it is the winner, i.e. $M_k = \sum_{s=1}^SI(W_s=k)$ (here $I$ denotes the indicator function). 

In the analyses below, $Z$ is considered to be a random variable. $V$, $W$, $P$ and $M$ are also random variables which depend on $Z$.

\subsection{District-wise Polarization Model (DPM)}
First of all, we consider the case where in each district the voters choose a party, based on local popularity. In other words, if $n_{sk}$ electors in district $s$ have already expressed support for party $k$, a new elector in that district will choose $k$ based on $n_{sk}$, but will also account for its country-wide popularity $\theta_k$. This model is a realistic representation of the voting behavior in many countries, where people often make a trade-off between the local candidate and the top leadership of a party before choosing to vote for it.

This model is based on the famous Chinese Restaurant Process (CRP)~\cite{crp} which is a subset of the CRF considered in the next subsection. The problem with this model is that, there is no way to ensure that the total number of votes obtained by the different parties is fixed to some known value. The proportion $\theta$ will be maintained approximately. but not exactly. However, here we can constrain the total number of votes obtained by each party $(v_1,\dots,v_K)$, by deactivating the choice of each party once the number of votes it gets from the different districts, maintained by book-keeping variable $m_k$, reaches the stipulated value $v_k$. If we denote by $X_{si}$ the vote of the $i$-th voter in district $s$, then its distribution is as follows:
\begin{eqnarray}
prob(X_{si}=k) \propto (\gamma_sn_{sk}+(1-\gamma_s)\theta_k)I(m_{k}<v_k)
\end{eqnarray}

$\gamma_s$ is the polarization parameter specific to district $s$. A high value of $\gamma_s$ indicates that electors in that district tend to choose the locally popular party, with less influence of the overall popularity of the parties indicated by $\theta$, and it creates the possibility of diversity across the districts. If $\gamma_s$ is close to $0$ in all districts, then the proportion of votes will reflect $\theta$ everywhere, and almost all districts will have the same winner.

\subsection{Elector Community Model (ECM)}
This model is based Hierarchical Dirichlet Process (HDP)~\cite{hdp} for grouped data. The HDP first considers a measure $P$, which follows the stick-breaking or GEM distribution. Next, for every data group $i$, a measure $Q_i$ is created from $P$ using a stick-breaking process. Finally, $n_i$ samples are drawn from $Q_i$, as the data-points associated with the group $i$. In this case, we can identify each group as a district, and $n_s$ as the number of electors in district $s$. The base distribution $H$ can be considered as the overall vote share $\theta$, and $Q_s$ is the vote share of the $K$ parties specific to the district $s$. Accordingly $n_s$ votes are polled for the different parties, as $\{V_{s1},\dots,V_{sK}\}$ by sampling from the distribution $Q_s$, and the winners are calculated. 

This model becomes more interesting and suitable for the voting scenario when we consider the Chinese Restaurant Franchise (CRF) representation of the HDP, which is obtained by marginalizing over $P$ and $Q$. This representation shows that the ``customers" (i.e. data-points) within each group arrange themselves in communities (clusters), by choosing ``tables" in an imaginary Chinese Restaurant, according to a rule indicated in Equation~\ref{eq:crf1}. Then a ``dish" is served to each table, across all the restaurants (i.e. groups), according to Equations~\ref{eq:crf2}. 

In our setting, the electors within each district first form communities among themselves (which we denote by $C$), and then all the members of a community vote from the same party (denoted by $D$). The process of community-formation is analogous to intra-district polarization among the electors. This is a common feature in the elections of many countries, as people vote according to the influence of their social communities rather than by individual choice. The communities are not uniformly sized, rather there are a few big and many small communities, due to the self-reinforcing (``rich getting richer") nature of Equation~\ref{eq:crf1}. The cluster members choose a party to vote for, taking into account the voting process across other districts. Each community tends to vote for a party which is already popular in other communities. This is also a realistic feature of elections in many countries, where people have a tendency to vote for that party whom they consider the strongest.

\begin{eqnarray}\label{eq:crf1}
prob(C_{si} = j) &=& \frac{n_{sj}}{i-1+\alpha_s} \textit{ if $n_{sj}$ electors have already joined community $j$ } \nonumber \\
                 &=& \frac{\alpha_s}{i-1+\alpha_s} \textit{ if $j$ is a new community}
\end{eqnarray}
\begin{eqnarray}\label{eq:crf2}
prob(D_{c} = k) &=& \frac{v_k+\beta.\theta}{c-1+\beta} \textit{ if $v_k$ communities of electors have already voted for party $k$ } \nonumber \\
\end{eqnarray}
Here, $\alpha$ and $\beta$ are two parameters of the model. High value of $\alpha$ increases the polarization within each district, while high value of $\beta$ creates high polarization \emph{across districts}, a situation where a small number of parties account for most of the votes. Once again, to make sure the parties get votes according to pre-specified $\theta$, we modify the second process (Equation~\ref{eq:crf2}) so that we can keep a record of the number of votes obtained by each party as the process proceeds. Once the party accumulates the votes stipulated for it, the process ensures that it gets no further votes.

\subsection{Party-wise Concentration Model (PCM)}
%Now we consider a model where electors can move and register themselves in districts. In doing so, each elector prefers districts which already have many electors who favour the same party. 
Now we consider a model where each party can choose to distribute it support across the districts. According to our assumption, each party has the capacity to win a certain number of votes, based on its popularity proportion. Each party can campaign in such a way that either its votes get concentrated in a few districts, or it can be spread across districts. 

The effect of this model is to create local concentrations of support in favour of different parties. Such concentration of support is necessary for parties to be effective in district-based elections. It is also a realistic phenomena, because support to political party is often based on social identities, and in most countries people choose residential areas based on social identities. For this model, we once again use the Chinese Restaurant Process model as in the District-wise Polarization Model. But this time we make the process two-step: each person is first assigned to a party, then (s)he is assigned a district based on concentration of support for that party. This model conflates party allegiance and social identity, without modeling them explicitly. Once again, we make sure that the total number of votes obtained by each party and the capacity of each district is maintained.
\begin{eqnarray}
prob(X_i=k) &\propto& \theta_{k}I(m_k<v_k) \nonumber \\
prob(Z_i=s|X_i=k) &\propto& (\eta_kV_{sk}+(1-\eta_k) U(\{1,\dots,K\})I(l_s<n_s)
\end{eqnarray}
Here, $m_k$ and $l_s$ are book-keeping variables, keeping track of the number of votes for party $k$ and number of persons in district $s$ at any point.

Like the parameter $\gamma_s$ in the previous model, high value of the parameter $\eta_k$ encourages voters of party $k$ to concentrate in a few districts, instead of spreading out uniformly. If all parties have low value of $\eta_k$, then once again the vote distribution in all districts will mirror $\theta$, and the most popular party overall will win all seats. Concentration of votes is particularly beneficial to parties which are less popular overall, it allows them to create local strongholds where they can win, even if they are non-existent elsewhere.

%\section{Theoretical Results: Effects of Concentration}

%\subsection{Bounds on the Number of Seats}

%\begin{itemize}
%    \item In a two-party setting, what is the maximum number of seats that any party can win, if the other party has the executive power to decide the number and/or size of the districts? (assuming voters can move between districts)
%    \item In a multi-party setting, what is the maximum number of seats that any arbitrary party can win, provided that the number and/or size of districts is fixed?
%    \item Same as above, if any other party has the power to set the number and/or size of districts?
%    \item Given the vote-share of all parties, what is the most competitive result achievable (defined in terms of some measure of dispersion between the seats won)
%    \item In a game-theoretic setting where voters from both parties can relocate, what are the equilibriums achievable (assuming district sizes fixed)?
%\end{itemize}

\section{Parameter Estimation by Approximate Bayesian Computation}

Clearly, these models have many parameters. If we wish to explain and analyze the results of actual elections using the models, we need to estimate these parameters. For parameter estimation, maximum-likelihood and Bayesian approaches are well-known. But in case of our models it is intractable to get a closed-form expression of the likelihood function, especially because of the book-keeping variables. So we look to likelihood-free inference techniques using Approximate Bayesian Computation (ABC).

\subsection{Summary Statistics}
For most ABC approaches, we need a low-dimensional representation of the data-points using which we wish to compute the posterior approximately. One well-known way of getting such a low-dimensional representation is by using summary statistics defined by the user. Various studies in scientific disciplines such as Ecology and Physics use ABC approaches by defining such statistics specific to the particular problems they address. In this case, we too define the following summary statistics, which can be easily calculated from $\{V_{s1},\dots,V_{sK}\}$, i.e. the number of votes obtained by each party in each district.

The summary statistics we propose are as follows:
\begin{enumerate}
    \item Number of districts ``won" by each party according to voting rule ($K$-dim)
    \item The mean fraction of votes won by each party across all districts ($K$-dim) 
    \item The standard deviation of the fraction of votes won by each party across all districts ($K$-dim)
    \item The standard deviation of the number of votes won by the parties in each district ($S$-dim)
    \item The mean and standard deviation of the winning margin across all districts ($2$-dim)
\end{enumerate}
Thus, the size of the summary statistics proposed here is $(3K+S+2)$, significantly less than the size $SK$ of the full data $V$.

\subsection{ABC Rejection Algorithm}
The simplest approach for this problem of parameter estimation is the ABC Rejection algorithm~\cite{abcrejection,lratio,logistic}. Here we sample the model parameters, denoted by $\psi$ from a suitable prior distribution. Using each sampled paramater value, we run the simulation and get the result $x$, from which we calculate the summary  statistics $S(x)$. Next, we compare $S(x)$ with the summary statistics $S(x_0)$ computed from the observed value $x_0$. If they are close enough to each other, then we accept the sample of the parameters, otherwise we reject it. The process is repeated until we have a large enough set of samples. Using this set we can compute the posterior distribution of the parameters by histogram analysis, and the mode of this posterior forms the Maximum a-Posteriori (MAP) estimate $\psi_{MAP}$. On the other hand, we can find that sample which creates the simulation summary statistics which is closest to the observed data, and use it as the optimal estimate $\psi_{OPT}$. %This is indicated as Algorithm 1.

%\begin{algorithm}[H]
%\SetAlgoLined
%\KwResult{Output: Posterior distribution and point estimate of the parameters}
% $ACC=\{\}$\;\\
% \While{$|ACC| < maxsize$}{\\
%  Sample $\psi \sim P$ (prior)\;\\
%  $x = F(\psi)$ (simulation)\;\\
%  \If{$||S(x)-S(x_0)||< \epsilon$}{ (compare summary statistics)\;\\
%   $ACC=ACC\cup\psi$ (accept sample)\;\\
%  }
%}
%   $Q = histogram(ACC)$ (compute posterior)\;\\
%   $\Psi_{MAP}=\arg\max Q$\;\\
%   $\Psi_{OPT}=\underset{{x\in ACC}}\arg\min{||S(x)-S(x_0)||}$
% \caption{ABC Rejection Algorithm}
%\end{algorithm}
%
%
%\begin{algorithm}[H]
%\SetAlgoLined
%\KwResult{Output: Posterior distribution and point estimate of the parameters}
%  $SEED=\{\}$\;\\
% \While{$n$ iterations}{
%  Sample $\psi \sim P$ (prior)\;\\
%  SEED=SEED$\cup\psi$\;\\
%  }
% SEEDERR = $\{||S(x)-S(x_0)||: x \in SEED \}$\;\\
% NEWSEED = $\{x : x \in SEED, SEEDERR(x) < percentile(SEEDERR,p)\}$\;\\
% C=covariance matrix from NEWSEED\;\\
% $ACC = \{\}$\;\\
% \While{$|ACC| < maxsize$}{\\
%  Uniformly sample $y$ from NEWSEED \;\\
%  Sample $\psi \sim \mathcal{N}(y,C)$ \;\\
%  $x = F(\psi)$ (simulation)\;\\
%  \If{$||S(x)-S(x_0)||< \epsilon$}{ \;\\
%   $ACC=ACC\cup\psi$ (accept sample)\;\\
%  }
%}
% $Q = histogram(ACC)$ (compute posterior)\;\\
% $\Psi_{MAP}=\arg\max Q$\;\\
% $\Psi_{OPT}=\underset{{x\in ACC}}\arg\min{||S(x)-S(x_0)||}$
% \caption{ABC Explore-Exploit Rejection Algorithm}
%\end{algorithm}

The big problem with this approach is that it is very slow, as most samples are likely to be rejected. That is why, once a sample is accepted, we may consider searching in the neighborhood of the accepted sample rather than sampling again from the prior. But then we may end up getting stuck at a local optima in the parameter space. So we use the explore-exploit approach. Here we first draw a limited number of samples from the prior and choose the best few among them as seeds (explore phase), and then we draw more samples around them, by using Gaussian distribution (exploit phase). Like Algorithm 1, we now accept those samples for which the simulation summary statistics are are close enough to $S(x_0)$. On the basis of these samples, we can calculate $\psi_{MAP}$ and $\psi_{OPT}$. We call this as Algorithm 2 or ABC Explore-Exploit Rejection, which is a modified version of SLAM algorithm~\cite{slam}.

\section{Regression-based approach to Likelihood-free Inference}
Since rejection-based ABC algorithms are slow, there have been attempts to employ supervised learning (classification/regression) to find the optimal solutions. Here, we try out two approaches to estimate the model parameters by posing this as a regression problem (multivariate regression in case of ECM and PCM) - logistic regression and neural networks. 

For this purpose, we first generate a training set by drawing $N$ samples of the model parameters $\{y_i\}_{i=1}^N$, and also the popularity proportion $\theta$ of the parties, the number of electors $N$ and number of districts $S$. Once we have a fairly diverse but representative set of samples, we run the models on them, to get the outputs $V$. Using $N$, $S$, $\theta$, $V$ we construct the feature vectors $\{x_i\}_{i=1}^N$, along with the parameters as corresponding output values $\{y_i\}_{i=1}^N$. Using this training set, we are now ready for the regression.

\subsection{Logistic Regression}
Logistic regression is a method for probabilistic classification. In this case, we utilize it to solve our regression problem by using it in the form of the bisection method. Given any feature vector $x$ (simulation details) and a model parameter estimate $y'$, the aim is to predict whether the correct parameter value $y$ (by which the model can produce the result encoded in $x$ from the input encoded in $x$) is greater than $y'$ or not. Clearly this is a binary classification task, which can be done by logistic regression. Depending on the classifier's prediction, we shift $y'$ either above or below its current value, midway within the feasible region (for example $\gamma$ of DPM and $\eta$ of PCM must lie in $(0,1)$). The feasible region also shrinks at each step, as in a decision tree. We continue this process till $y'$ converges. For every test value $y'$, we train the logistic regression classifier on the training set by re-labeling all samples based on $sign(y-y')$. This is not a time-consuming process and can be done on-the-fly. If the model has multiple parameters (ECM and PCM), we do this separately for all parameters. It is a simplistic but efficient and interpretable approach.

\subsection{Neural Networks}
The next approach to regression which we consider is Deep Neural Network. In this case we do not need the bisection method as the neural network itself can produce real vector outputs. We tried with several architectures, and found the optimal one to have two dense layers with 33 and 38 hidden units (when the input is 24-dimensional along with feature scaling) along with $20\%$ drop-out and Scaled Exponential Linear Unit (SELU) and Adam optimizer.

\subsection{Regression-Rejection Hybrid Approach}
We tried both the approaches mentioned above (Logistic Regression and Neural Network) on the real-world data mentioned in Section 7. Unfortunately, neither of the above approaches could produce optimal values of the desired parameters. Though they came quite close to the optimal values, in these models even a small change in the parameter values can have an effect on the results, as demonstrated on synthetic data in Section 6.

Hence, we advocate a hybrid approach, which combines both regression and rejection. We propose to replace the ``explore" phase of Algorithm 2 with the regression-based approach, to get an estimate called $SEED$. The covariance matrix around $SEED$ can be provided by the user. The exact value of the parameters can then be found by sampling around $SEED$. %The algorithm is introduced formally as Algorithm 3.

%\begin{algorithm}[H]
%\SetAlgoLined
%\KwResult{Output: Posterior distribution and point estimate of the parameters}
% $SEED= f(S(x_0))$   (f: pre-trained regression function)\;\\
% $ACC = \{\}$\;\\
% \While{$|ACC| < maxsize$}{\\
%  Sample $\psi \sim \mathcal{N}(SEED,C)$ ($C$: covariance matrix supplied by user)\;\\ 
%  $x = F(\psi)$ (simulation)\;\\
%  \If{$||S(x)-S(x_0)||< \epsilon$}{ \;\\
%   $ACC=ACC\cup\psi$ (accept sample)\;\\
%  }
%}
% $Q = histogram(ACC)$ (compute posterior)\;\\
% $\Psi_{MAP}=\arg\max Q$\;\\
% $\Psi_{OPT}=\underset{{x\in ACC}}\arg\min{||S(x)-S(x_0)||}$
% \caption{Hybrid Regression-Rejection Algorithm}
%\end{algorithm}

\section{Analysis of Models through Simulation}
In this section, we illustrate through simulations various aspects of our models on synthetic data. We consider a simple 2-party system. Keeping the number of districts $S=100$ and number of electors $N=1000000$, we vary the popularity proportion $\theta$, and observe the results under different settings of the parameters of our models. For each setting, we carry out 100 simulations and report: i) The mean number of seats won by the parties, and ii) The mean and variance of winning margins.

\begin{table}
    \centering
    \begin{tabular}{|c|c|c|c|c||c|c|c|c|}
    \hline
                    & \multicolumn{4}{c||}{Number of seats won by Party 1} & \multicolumn{4}{c|}{St. Dev. of votes for Party 1} \\
    \hline
    $\theta_1$      & 0.5 & 0.6 & 0.7 & 0.8 & 0.5 & 0.6 & 0.7 & 0.8 \\
    \hline
    $\gamma=0.25$   & $50\pm3$ & 100      & 100      & 100      & 0.01 & 0.01 & 0.01 & 0.01\\
    $\gamma=0.50$   & $50\pm3$ & 100      & 100      & 100      & 0.01 & 0.01 & 0.01 & 0.01\\
    $\gamma=0.75$   & $50\pm3$ & $96\pm2$ & 100      & 100      & 0.05 & 0.06 & 0.06 & 0.07\\
    $\gamma=0.90$   & $50\pm3$ & $72\pm3$ & $83\pm3$ & $93\pm2$ & 0.15 & 0.16 & 0.18 & 0.18\\
    $\gamma=0.99$   & $50\pm3$ & $60\pm3$ & $70\pm2$ & $80\pm2$ & 0.28 & 0.30 & 0.31 & 0.29\\
    \hline
    \end{tabular}
    \caption{Synthetic results in a two-party election under different parameter settings (shown in the rows) under District-wise Polarization Model, showing number of seats won by party 1 (whose popularity is indicated for each column) and the standard deviation of its votes across the districts.}
    \label{tab:sim1_1}
\end{table}

\textbf{Observations:District-wise Polarization Model}
\begin{enumerate}
    \item Increase of polarization usually helps the less popular party to increase its seat share
    \item Increase of polarization increases the mean margin of victory in the districts (not shown in the table)
    \item Increase of polarization increases the variance in the number of votes obtained by it across the districts
\end{enumerate}

\begin{table}
    \centering
    \begin{tabular}{|c|c|c|c|c||c|c|c|c|}
    \hline
                    & \multicolumn{4}{c||}{Number of seats won by Party 1} & \multicolumn{4}{c|}{Mean Winning Margin} \\
    \hline
    $\theta_1$                  & 0.5 & 0.6 & 0.7 & 0.8 & 0.5 & 0.6 & 0.7 & 0.8 \\
    \hline
    $\{\alpha=1,\beta=0.25\}$   & $50\pm4$ & $62\pm5$ & $73\pm4$ & $84\pm3$  & 0.80 & 0.82 & 0.84 & 0.88\\
    $\{\alpha=1,\beta=0.75\}$   & $51\pm4$ & $63\pm5$ & $71\pm4$ & $80\pm3$  & 0.83 & 0.84 & 0.85 & 0.89\\
    $\{\alpha=1,\beta=0.99\}$   & $51\pm3$ & $61\pm3$ & $70\pm2$ & $79\pm2$  & 0.89 & 0.90 & 0.91 & 0.93\\
    $\{\alpha=5,\beta=0.25\}$   & $50\pm4$ & $72\pm4$ & $89\pm3$ & $97\pm2$  & 0.64 & 0.66 & 0.72 & 0.80\\
    $\{\alpha=5,\beta=0.75\}$   & $48\pm7$ & $69\pm8$ & $83\pm6$ & $94\pm4$  & 0.68 & 0.69 & 0.75 & 0.81\\
    $\{\alpha=5,\beta=0.99\}$   & $49\pm9$ & $64\pm9$ & $70\pm9$ & $84\pm8$  & 0.69 & 0.85 & 0.88 & 0.89\\
    $\{\alpha=10,\beta=0.25\}$  & $51\pm5$ & $80\pm4$ & $95\pm2$ & 100     & 0.60 & 0.63 & 0.70 & 0.80\\
    $\{\alpha=10,\beta=0.75\}$  & $47\pm12$& $75\pm11$ & $90\pm5$ & $96\pm4$ & 0.65 & 0.66 & 0.72 & 0.80\\
    $\{\alpha=10,\beta=0.99\}$  & $50\pm12$& $66\pm12$ & $73\pm14$& $84\pm10$& 0.76 & 0.82 & 0.84 & 0.88\\
    \hline
    \end{tabular}
    \caption{Synthetic results in a two-party election under different parameter settings (shown in the rows) under Elector Community Model, showing number of seats won by party 1 (whose popularity is indicated for each column) and the average winning margin.}
    \label{tab:sim2_1}
\end{table}

\textbf{Observations:Elector Community Model}
\begin{enumerate}
    \item Increase in the community parameter $\alpha$ creates smaller communities, which benefits the more popular party
    \item Increase in the concentration parameter $\beta$ benefits the party that is less popular.
    \item High values of both parameters increase uncertainty of the results
    \item Low values of $\alpha$, forms large communities and increases margins of victory
\end{enumerate}

\begin{table}
    \centering
    \begin{tabular}{|c|c|c|c|c||c|c|c|c|}
    \hline
                    & \multicolumn{4}{c||}{Number of seats won by Party 1} & \multicolumn{4}{c|}{Mean Winning Margin} \\
    \hline
    $\theta_1$                    & 0.5 & 0.6 & 0.7 & 0.8 & 0.5 & 0.6 & 0.7 & 0.8 \\
    \hline
    $\{\eta_1=0.5,\eta_2=0.5\}$   & $51\pm3$ & 100     & 100       & 100   & 0.51 & 0.60 & 0.70 & 0.80 \\
    $\{\eta_1=0.5,\eta_2=0.7\}$   & $55\pm5$ & 100     & 100       & 100   & 0.52 & 0.60 & 0.70 & 0.80 \\
    $\{\eta_1=0.5,\eta_2=0.99\}$  & $50\pm3$ & $71\pm2$& $90\pm2$  & 99    & 0.63 & 0.65 & 0.72 & 0.80 \\
    $\{\eta_1=0.7,\eta_2=0.5\}$   & $46\pm3$ & 100     & 100       & 100   & 0.53 & 0.60 & 0.70 & 0.80 \\
    $\{\eta_1=0.7,\eta_2=0.7\}$   & $50\pm4$ & $99\pm1$& 100       & 100   & 0.53 & 0.60 & 0.70 & 0.80 \\
    $\{\eta_1=0.7,\eta_2=0.99\}$  & $49\pm4$ & $71\pm3$& $89\pm2$  & $98\pm1$& 0.63 & 0.66 & 0.71 & 0.80 \\
    $\{\eta_1=0.99,\eta_2=0.5\}$  & $51\pm2$ & $75\pm2$& $94\pm1$  & 100   & 0.63 & 0.64 & 0.71 & 0.80 \\
    $\{\eta_1=0.99,\eta_2=0.7\}$  & $51\pm2$ & $77\pm3$& $91\pm2$  & 100   & 0.63 & 0.65 & 0.71 & 0.80 \\
    $\{\eta_1=0.99,\eta_2=0.99\}$ & $48\pm2$ & $64\pm2$& $81\pm2$  & $92\pm1$& 0.72 & 0.73 & 0.76 & 0.82 \\
    \hline
    \end{tabular}
    \caption{Synthetic results in a two-party election under different parameter settings (shown in the rows) under Partywise Concentration Model, showing number of seats won by party 1 whose popularity is indicated for each column, and also the winning margins}
    \label{tab:sim3_1}
\end{table}

\textbf{Observations: Partywise Concentration Model} 
\begin{enumerate}
    \item If both parties have nearly equal popularity, a party should have either very low or very high concentration relative to the other, to maximize its seat share.
    \item If there is a significant gap in popularity between the parties, increasing concentration of support is good for the less popular party, but bad for the more popular party.
    \item Increase in concentration of support for any party increases the variance in the number of votes obtained by it across the districts (not shown in the table)
\end{enumerate}

The results presented show that the part which is more popular overall, usually has an extra advantage in terms of seats won, i.e. the proportion of seats won by them usually exceeds their proportion of popularity (vote share). However, it is possible for the smaller party to offset this disadvantage in certain circumstances. In case of DPM, high value of $\gamma$ causes the supporters to concentrate in space, and the seat-share reflects the vote share. In case of ECM, we see that as $\alpha$ and $\beta$ both increase, the less popular party starts punching above its weight, i.e. its seat-share approaches and may even exceed its vote-share.

\section{Analysis of Indian Elections}

\begin{table}[]
    \centering
    \begin{tabular}{|c|c|c|c|c|c|}
      \hline
        & Party A & Party B & Party C & MWM & SWM\\ 
      \hline
       2013$\theta$                    & 0.30 & 0.33  & 0.25 & NA & NA\\
       Proportional                    & 21 & 23 & 18 &   & \\
       \hline   
       SPM($\gamma=0.89$)              & \textbf{25} & \textbf{35} & 10 & 0.37 & 0.07\\
       SPM($\gamma=0.99$)              & 21 & \textbf{24} & 18 & 0.55 & 0.18\\
       \hline
       ECM($\{\alpha=16,\beta=0.24\}$) & \textbf{28} & \textbf{34} &  8 & 0.39 & 0.07\\
       ECM($\{\alpha=1,\beta=0.99\}$)  & 21 & 23 & 18 & 0.77 & 0.19\\
       \hline       
       EMM($\eta=\{0.55,0.89,0.84\}$)  & \textbf{27} & \textbf{34} & 9  & 0.36 & 0.09\\
       EMM($\eta=\{0.99,0.99,0.99\}$)  & \textbf{24} & \textbf{28} & 18 & 0.51 & 0.11\\
       \hline       
       Actual                          & \textbf{28} & \textbf{34} & 8 & 0.39 & 0.06\\
      \hline
      \hline
       2015$\theta$                    & 0.54 & 0.32 & 0.10 & NA & NA\\
       Proportional                    & 38   & 22   & 7    & NA & NA\\
       \hline       
       DPM($\gamma=0.86$)              & \textbf{68} & 2  & 0 & 0.54 & 0.09\\ 
       DPM($\gamma=0.99$)              & 37 & 23 & 4 & 0.71 & 0.21\\
       \hline       
       ECM$(\{\alpha=30,\beta=0.21\})$ & \textbf{67} &  3 & 0 & 0.55 & 0.07\\
       ECM$(\{\alpha=1,\beta=0.99\})$  & 38 & 23 & 7 & 0.87 & 0.17\\
       \hline       
       PCM($\eta=\{0.74,0.89,0.68\}$   & \textbf{67} & 3  & 0 & 0.54 & 0.05\\
       PCM($\eta=\{0.99,0.99,0.99\}$   & \textbf{49} & 19 & 2 & 0.64 & 0.14\\
       \hline       
       Actual                          & \textbf{67} & 3  & 0 & 0.55 & 0.07\\
       \hline
       \hline        
       2020$\theta$                    & 0.54 & 0.39 & 0.05 & NA & NA\\
       Proportional                    & 37   & 27   & 4    & NA & NA\\
       \hline 
       DPM($\gamma=0.87$)              & \textbf{60} & 10 & 0 & 0.55 & 0.08\\
       DPM($\gamma=0.99$)              & 38 & 28 & 4 & 0.73 & 0.18\\
       \hline   
       ECM($\{\alpha=36,\beta=0.57\}$) & \textbf{62} &  8 & 0 & 0.55 & 0.06\\ 
       ECM($\{\alpha=1,\beta=0.99\}$)  & \textbf{40} & 25 & 4 & 0.85 & 0.17\\
       \hline       
       PCM($\eta=\{0.72,0.80,0.72\}$)  & \textbf{62} &  8 & 0 & 0.55 & 0.08\\
       PCM($\eta=\{0.99,0.99.0.99\}$)  & \textbf{43} & 27 & 0 & 0.66 & 0.13 \\       
       \hline       
       Actual                          & \textbf{62} &  8 & 0 & 0.55 & 0.06\\
       \hline
    \end{tabular}
    \caption{Elections in Delhi-NCR, India: The actual and model-predicted performances of 3 top parties in past 3 assembly elections (2013, 2015, 2020), based on their popularity proportions (vote share). For each model, results are shown with the default parameters as well as optimal settings as computed by Hybrid  Regression-Rejection Algorithm. In each election, the number of seats won by each party under various settings is compared with the seats proportional to their vote share, and the cases where a party gains seats are highlighted}
    \label{tab:sim4}
\end{table}

Finally, we attempt to understand actual elections held in India over the last 10 years. We use the hybrid regression-rejection algorithm discussed earlier to find the optimal parameters to fit each model to each election. We also explore alternative results to these elections, under the same proportion of popularity, but different parameter settings. Due to the volatile nature of the subject, we anonymize the political parties.

%\subsection{Small Election: Delhi Assembly}
We consider the Indian province of Delhi National Capital Region (NCR), whose local governing body has 70 seats. Around 9 million people (on average) participate in the elections, with roughly equal distribution of electors across the 70 districts. Since 2013, 5 elections have taken (local and national) in which 3 main political parties have competed, along with small parties and independent candidates. The overall vote-shares (popularity proportions) of the parties have varied across the elections.

In the Table 3, we show the expected results according to our models, and compare them with the actual results for the local assembly elections of 2013, 2015 and 2020. Due to lack of space, only two parameter settings are shown: default and optimal. Optimal setting is the one which is estimated using the Approximate Bayesian technique discussed in the previous section. The names of the parties have been anonymized. In Table 4, we use the optimal parameter settings estimated in Table 3 for the local elections to extrapolate the national election results of 2014 and 2019.

\begin{table}[]
    \centering
    \begin{tabular}{|c|c|c|c|c|c|}
      \hline
        & Party A & Party B & Party C & MWM & SWM\\ 
      \hline
       2014$\theta$               & 0.33 & 0.46 & 0.15 & NA & NA\\
       Proportional               & 23   & 32   & 11   &    & \\
       \hline       
       DPM(estimated)             & $10\pm4$ & \textbf{60}$\pm4$ & 0 & 0.47 & 0.09\\ 
       DPM($\gamma=0.99$)         & 24 & 33 & 10 & 0.65 & 0.2\\
       \hline  
       ECM(estimated)                  & $13\pm3$ & \textbf{57}$\pm3$ & 0 & 0.48 & 0.08\\
       ECM($\{\alpha=1,\beta=0.99\}$)  & 23 & 31 & 12 & 0.85 & 0.18\\
       \hline       
       PCM(estimated)                  & $8\pm3$ & \textbf{62}$\pm3$ & 0  & 0.47 & 0.1\\
       PCM($\eta=\{0.99,0.99,0.99\}$)  & 23 & \textbf{43} & 4  & 0.58 & 0.13\\
       \hline       
       Actual                     & 10 & \textbf{60} & 0 & & \\
       \hline
       \hline        
       2019$\theta$               & 0.18 & 0.56 & 0.23 & NA & NA\\
       Proportional               & 13   & 39   & 16   &    &  \\
       \hline       
       DPM(estimated)             &  1 & \textbf{68} & 1  & 0.56 & 0.11\\
       DPM($\gamma=0.99$)         & 12 & 41 & 16 & 0.70 & 0.22\\
       \hline  
       ECM(estimated)                  &  0 & \textbf{69} & 1  & 0.56 & 0.08\\
       ECM($\{\alpha=1,\beta=0.99\}$)  & 12 & 39 & 17 & 0.83 & 0.17\\
       \hline       
       PCM($\eta=\{0.7,0.95.0.9\}$)    & 0  & \textbf{70} & 0  & 0.56 & 0.11 \\
       PCM($\eta=\{0.99,0.99.0.99\}$)  & 6  & \textbf{52} & 12 & 0.62 & 0.15 \\
       \hline       
       Actual                     & 0 & \textbf{65} & 5  & & \\
       \hline
       \hline       
    \end{tabular}
    \caption{Elections in Delhi-NCR, India: The actual and model-predicted performances of 3 top parties in past 2 parliamentary elections (2014 and 2019), based on their popularity proportions (vote share). For each model, results are shown with the default parameters as well as the parameters estimated by ABC Explore-Exploit Rejection algorithm from the 3 assembly elections, as shown in the previous table. In each election, the number of seats won by each party under various settings is compared with the seats proportional to their vote share, and the cases where a party gains seats are highlighted}
    \label{tab:sim4}
\end{table}

The results show that Elector Community Model is most successful in reproducing the actual results, closely followed by the Partywise Concentration Model. It is also seen that the results can be considerably different under other parameter settings, such as the default settings used in each case. In Table 4, it is interesting to note that although all the models could roughly predict the results for the 2014 national elections using the optimal parameters estimated from the three assembly elections, none could do so for the 2019 national elections.

\section{Conclusion}
In this paper, we explored the effects of spatial distribution of electors on the results of district-based elections. We considered 3 stochastic models for such spatial distribution, based on the observations that i) an individual's vote is influenced by overall popularity of parties, ii) people tend to vote for parties based on their social identity/community affiliation, and iii) people with similar social identity tend to live geographically close. We saw that these factors allow political parties to establish certain districts as their strongholds, and thus they may be able to win disproportionately more seats in the governing body, even if their popularity is less in the overall population. We also explored the parameter estimation problem for the proposed models, and for this we designed algorithms based on Approximate Bayesian Computation as well as regression. In future work, we would like to i) consider more sophisticated and realistic models of spatial distribution of electors, ii) use this framework to study more elections in various countries which have district-based voting system, iii) collaborate with psephologists to understand how the parameters vary from one election to another.

\section{Acknowledgements}
The author thanks Vishesh Kumar and Shukur Ali, both undergraduate students in Indian Institute of Technology Kharagpur, for the deep neural network-based regression analysis.

\end{document}